\documentstyle[prd,tighten,aps]{revtex}
 
\newcommand{\be}{\begin{equation}}
\newcommand{\ee}{\end{equation}}
\newcommand{\bn}{\begin{eqnarray}}
\newcommand{\en}{\end{eqnarray}}
\newcommand{\p}{\partial}

\def\l{\label}

\begin{document}
\draft

\twocolumn[\hsize\textwidth\columnwidth\hsize\csname
@twocolumnfalse\endcsname

\title{Dual projection in a two-dimensional curved expanding universe}
\author{Everton M. C. Abreu$^a$}
\address{Departamento de Campos e Part\'{\i}culas, 
Centro Brasileiro de Pesquisas F\'{\i}sicas, \\[0pt]
Rua Xavier Sigaud 150, Urca, 22290-180, 
Rio de Janeiro, RJ, Brazil. \\[0pt]
${}^{a}$ Present address: Departamento de F\'{\i}sica Te\'orica, Universidade do Estado do Rio de Janeiro,\\
Rua S\~ao Francisco Xavier 524, Maracan\~a, 20550-013, Rio de Janeiro, Brazil\\
{\sf E-mail: everton@cbpf.br}}

\date{\today}
\maketitle

\begin{abstract}
\noindent  It is a well known result that the scalar field is composed of two chiral particles (Floreanini-Jackiw particles) of opposite chiralities.  Also, that a Siegel particle spectrum is formed by a nonmover field (a Hull's noton) and a FJ particle.  In this work we show that in a scalar field spectrum, in a curved expanding universe scenario, we  find a different result, two dynamical chiral fields.
\end{abstract}

\vskip2pc]

\newpage

\section{Introduction}

The study of chiral bosons in a $D=2$ flat space has a wide interest. They
occur as basic ingredients in the formulation of string theories and in a
number of two-dimensional statistical systems, like the fractional quantum
Hall effect phenomenology.  More recently, the M-theory can be achieved
by treating the chiral sectors in a more independent way. In superior
dimensions, the six-dimensional chiral bosons belong to the supergravity and
tensor multiplets in $N=1$, $D=6$ supergravity theories. They are necessary
to complete the $N=2$, $D=6$ supermultiplet of the $M$-theory five-brane.
Finally, a ten-dimensional chiral bosons appears in $IIB$, $D=10$
supergravity.

The dual projection technique \cite{wotzasek,aw,baw,baw2,eu}, strictly related
to canonical transformations \cite{bg}, has been presented initially in the
study of electromagnetic duality groups \cite{wotzasek}. The difference
between both concepts is that the dual projection is performed at the level
of the actions whereas a canonical transformation is performed at the
Hamiltonian level. However, as in this work we will consider only
first-order actions, the equivalence between the dual projection and the
canonical transformations becomes manifest \cite{bg}. The most useful and
interesting point in this dual projection procedure is that it is not based
on evidently even-dimensional concepts and may be extended to the
odd-dimensional situation \cite{baw} and also in non-Abelian systems \cite{bw}. 
In this work we will show that besides
the use of dual projection in even and odd dimensions in flat space, we can
obtain interesting results in curved space.

The noton is a nonmover field at the classical level \cite{hull}, carrying
the representation of the Siegel symmetry \cite{siegel}, that acquires
dynamics upon quantization. As another feature, it is a well known fact that
coupling chiral particles to external gravitational fields reveals the
presence of notons \cite{dgr}. At the quantum level in flat space, it was
shown \cite{aw} that its dynamics is fully responsible for the Siegel
anomaly. In other words, we can say that the importance of the inclusion of
a normalized external noton - the Hull mechanism - is to cancel the Siegel
anomaly \cite{hull} conveniently.

On the other hand, later than Siegel, Floreanini and Jackiw (FJ) introduced
an action which describes a free chiral boson in two dimensions \cite{fj},
but this action have some symmetry problems. In \cite{aw} it was shown that,
in fact both models (Siegel and FJ) for the chiral bosons are related by a
noton particle.

In this work we want to analyze the effect of the dual projection in a
two-dimensional curved space (a Friedmann space) with a metric that describes an expanding
universe model \cite{parker}.  We shall consider here the universe as a
curved space.  Notice that the main idea here is to analyze the different spectra found.

Quantum field theory in curved spacetime is an effective theory which is
able to make pretty reliable quantum gravity predictions in some specific
regimes. In a nutshell we can say that it investigates the consequences of
defining quantum fields on general background spacetime. Although it is
unable to describe nature in extreme regimes, as in the Planck scale, some
remarkable effects were already predicted in its context as, for instance,
that quantum effects must induce black holes to evaporate, in contrast to
the classical belief \cite{curved}.

Firstly we will show that the dual projection of a scalar field in a generic gravitational
background results in two notons.  This result is quite different from the curved space one, which shows two dynamical chiral fields interacting with the external gravitational field.   Notwithstanding, the final system is equivalent to a single dynamical non-chiral fields in the gravitational background, as it should be.  

The mentioned chirality separation is confirmed when we perform the flat space limit, bearing out the results obtained by Fronsdal {\it et al} in a series of papers on curved space \cite{fronsdal}, i.e., ``that a flat space result can be obtained from
the curved space one through a determined approximation''. 

The sequence of the paper is: in order to make this work self-consistent, in
the next section we review the $D=2$ dual projection applied to the scalar
field in a Minkowski space and to the Siegel chiral model, which
introduces the presence of the noton as in \cite{aw}. In Section 3 we 
perform the dual projection of a scalar field in a gravitational background with 
an arbitrary metric showing the presence of two notons.  And finally, we carry
out the dual projection of a two-dimensional curved expanding universe
analyzing its spectrum showing that at this time there is no notons. 
As usual, the conclusion and perspectives are discussed in the last section.

\section{The dual projection in a flat space}

The separation of a scalar field into its chiral components has been
introduced by Mandelstam \cite{mandelstam} in its seminal paper on 2D
bosonization. The chiral splitting for the non-Abelian side has been studied
by Polyakov and Wiegman \cite{pw}. However, the Abelian limit of
Polyakov-Wiegman decomposition does not coincide with Mandelstam's chiral
decomposition. The chiral separation in \cite{mandelstam} is based on a
first-order theory while that of \cite{pw} is second-order. Mandelstam's
chiral decomposition scheme can be obtained as a constraint over the theory
requiring the complete separability of the original action.

The dual projection in $D=2(2p+1)$ dimensions leads to a diagonal form of
the action (hence we can also call it as a diagonalization procedure). 
Here, differently from \cite{bw}, the two pieces manifest
completely unlike features: while one piece is chiral (a FJ particle), responsible for the
dynamical sector of the theory, the other carry the algebraic component (a noton particle), as shown in \cite{aw}. We select here only two examples, among others, in the
literature where the dual projection succeed.

{\bf A. The scalar field}.  In this section we will make a brief analysis of the dual projection of a
scalar field in a $D=2$ Minkowski space. This result was obtained a few
years back by Tseytlin and West \cite{tw}. Performing the inverse process,
confirming this result, the recent soldering formalism \cite{solda,adw}, 
showed us that the fusion of two FJ modes gives a scalar field. This makes
sense since a scalar field, in a $D=2$ flat space, has no chirality. Hence,
both FJ's particles interact each other so that as a result we have no
chirality at all. This example, well known, is very important to our
purpose, since we will see that in a curved space this behavior is not true
anymore.

In a few steps we can begin with the Lagrangian density of a scalar field, 
\be  \label{1}
{\cal L} = {\frac{1 }{2}}\, \partial_{\mu}\,\phi\,\partial^{\mu}\,\phi 
\,=\, {\frac{1 }{2}}\,{\dot{\phi}}^2\,-\,{\frac{1 }{2}}\,{\phi^{\prime}}^2,
\ee
where the metric is $(+,-)$.  Dot and prime means respectively the time
and space derivatives. Reducing the order of the time derivative, we can use
an auxiliary field (not necessarily the canonical momentum) so that we can
write 
\begin{equation}  \label{dois_1}
{\frac{1 }{2}}\,{\dot{\phi}}^2\,=\, \dot{\phi}\,p\,-\,{\frac{1 }{2}}\,p^2
\end{equation}
and substituting (\ref{dois_1}) in (\ref{1}) the Lagrangian reads, 
$
{\cal L}_1 \,=\, \dot{\phi}\,p\,-\,{\frac{1 }{2}}\,p^2 \,-\,{\frac{1 }{2}}\,{%
\phi^{\prime}}^2.
$
Performing the canonical transformations convenient constructed in order to
promote a complete field separation, we have, 
\begin{equation}  \label{4}
\phi\,=\, \rho\,+\,\sigma \qquad \mbox{and} \qquad
p\,=\,\rho^{\prime}\,-\,\sigma^{\prime}\;\;.
\end{equation}
Substituting it in ${\cal L}_1$ above, we have a new action, 
$
{\cal L}_2\,=\,\dot{\rho}\,{\rho^{\prime}}^2\,-\,{\rho^{\prime}}^2\,-\,\dot{%
\sigma}\,\sigma^{\prime}\,-\,{\sigma^{\prime}}^2,
$
which shows clearly that the spectrum of a scalar field is formed by two FJ
particles of opposite chirality as we said above. For an interested reader,
the relation between two chiral particles was discussed at the Hamiltonian
level in \cite{sonnenschein}.

There are indications that a deeper understanding of such issues as string
dynamics and fractional quantum Hall effect phenomenology can be achieved by
treating the chiral sectors in a more independent way. However, coupling
chiral fields to external gauge and gravitational fields is problematic. As
we said above, it was discussed in \cite{bw}, how the coupling of chiral
(Abelian) fields to external gravitational backgrounds can be achieved by
diagonalization (dual projection) of the first-order form of a covariant
scalar action. The theory reduces then to a sum of a left and a right FJ's
actions \cite{fj}, circumventing the problems caused by the lack of manifest
Lorentz invariance.

{\bf B. The Siegel chiral boson}.
Now we will present a small version of the results obtained in \cite{aw}. We
begin with the Siegel original classical Lagrangian density for a chiral
scalar field \cite{siegel} using the usual light-cone variables, 
$$
{\cal L}_{Siegel}=\partial_+\phi\,\partial_-\phi - \lambda_{++}(\partial_-%
\phi)^2  
={\frac{1 }{2}}\,\sqrt{-g}\,g^{\alpha\beta}\partial_{\alpha}\phi\,\partial_{\beta}\,\phi
$$
where the metric is 
$
g^{++}\,=\,0\:,\:g^{+-}\,=\,1\:,\:g^{--}\,=\,-\,2\,%
\lambda_{++}\:,
$
and we can say that the Lagrangian ${\cal L}_{Siegel}$ describes a lefton \cite{gs}.
A lefton (or a righton) is a particle which, besides to carry the dynamics
of the theory, it is liable for the symmetry of the system too. This
characterizes exactly a Siegel mode. Hence, it is different from a FJ's
particle \cite{fj} which carries only the dynamics of the system. Thereby, a
FJ particle can not be classified either as a lefton or a righton.

The symmetry content of the theory is well described by the Siegel algebra,
a truncate diffeomorphism that disappears at the quantum level. Hence ${\cal L}_{Siegel}$ is invariant under Siegel gauge symmetry which is an invariance under
the combined coordinate transformation and a Weyl rescaling of the form 
$
x^- \rightarrow \tilde{x}^-\,=\,x^-\,-\,\epsilon^- \quad \mbox{and} \quad
\delta g_{\alpha\beta} \,=\, -\,g_{\alpha\beta}\,\partial_-\,\epsilon^- .
$
The fields $\phi$ and $\lambda_{++}$ transform under these relations as
$
\delta \phi= \epsilon^-\,\partial_-\,\phi,$ and $  
\delta \lambda_{++}=
-\,\partial_+\,\epsilon^- +\epsilon^-\,\partial_-\,\lambda_{++}\,-\,
\partial_-\,\epsilon^-\,\lambda_{++}$,
and $\phi$ is invariant under the global axial transformation,
$
\phi \rightarrow \tilde{\phi}\,=\,\phi\,+\,\bar{\phi}
$.
It is easy to see that fixing the value of the multiplier as $\lambda_{++}=1$
in ${\cal L}_{Siegel}$ we can obtain the FJ model. As mentioned before, this was
considered, for a long time in the literature, the unique constraint between
these both representations of a chiral boson. The dual projection permit us
to see other intrinsic relations behind the model that can possibly (or not)
be hidden in its spectrum.

We will now begin to apply the dual projection procedure. This is done
introducing a dynamical redefinition in the phase space of the model. Using ${\cal L}_{Siegel}$ in Lorentz coordinates we can obtain the canonical momentum as 
\begin{equation}  \label{quatro}
\pi\,=\,\frac{\partial {\cal L}}{\partial \dot{\phi}}\,=\,\dot{\phi}%
\,-\,\lambda_{++}\,(\,\dot{\phi}\, -\,\phi^{\prime}\,)
\end{equation}
or, in other words 
\begin{equation}  \label{cinco}
\dot{\phi}\,=\,\frac{\pi\,-\,\lambda_{++}\,\phi^{\prime}}{1\,-\,\lambda_{++}}%
\;\;.
\end{equation}
After a little algebra, substituting (\ref{quatro}) and (\ref{cinco}) into ${\cal L}_{Siegel}$, the Lagrangian in the first-order form reads 
\begin{equation}  \label{cincoa}
{\cal L}_{Siegel}\,=\,\pi\,\dot{\phi}\,-\,\frac{{\phi^{\prime}}^2}{2}\,-\, {\frac{1 }{%
2}}\,\frac{(\,\pi\,-\,\lambda_{++}\,\phi^{\prime}\,)^2}{1\,-\,\lambda_{++}}%
\,-\, {\frac{\lambda_{++} }{2}}\,{\phi^{\prime}}^2 .
\end{equation}

As we said above, fixing the value of the multiplier as $\lambda
_{++}\rightarrow 1$ in (\ref{cincoa}) we get the FJ form. This value of $%
\lambda _{++}$ promotes a reduction of the phase space of the model to \cite%
{djt} 
$
\pi \rightarrow {\phi ^{\prime }} ,
$
and consequently the third term in (\ref{cincoa}) reduces to zero as $%
\lambda _{++}\rightarrow 1$. Therefore the dynamics of the system will be
described by a FJ action.

The above behavior in $\pi \rightarrow {\phi ^{\prime }} $ suggests the following canonical
transformations, analogous as in (\ref{4}): 
\begin{equation}
\phi \,=\,\varphi \,+\,\sigma \qquad \mbox{and}\qquad \pi \,=\,\varphi
^{\prime }\,-\,\sigma ^{\prime }\;\;,  \label{sete}
\end{equation}
and we stress that these fields are independent as they originate from
completely different actions. After substituting (\ref{sete}) into (\ref%
{cincoa}) to perform the dual projection we find a diagonalized Lagrangian, 
$
{\cal L}\,=\,\varphi ^{\prime }\,\dot{\varphi}\,-\,{\varphi ^{\prime }}%
^{2}\,-\,\sigma ^{\prime }\,\dot{\sigma}\,-\,\eta _{+}\,{\sigma ^{\prime }}%
^{2} ,  
$
where 
$
\eta _{+}\,=\,\frac{1\,+\lambda _{++}}{1\,-\lambda _{++}}\;\;. 
$
The effect of dual projection procedure into the first-order Siegel theory,
equation (\ref{cincoa}), was the creation of two different internal spaces
leading to the $Z_{2}$ group of dualities (a discrete group with two
elements) \cite{wotzasek,baw} and the other is the diffeomorphism group of
transformations. Clearly we can see that the chirality/duality group and the
symmetry group are in different sectors. The first is obviously a FJ mode
and the other is a noton mode. This result is complementary to the
established knowledge, where the FJ action is interpreted as a gauge fixed
Siegel action \cite{siegel}. Under this point of view, we look at the gauge
fixing process as the condition that sets the noton field to vanish. It can
be proved \cite{aw} that this noton is totally responsible for the
symmetries, both classically and quantically.

\section{Dual projection in curved space}

In this section we will use the dual projection to analyze the spectrum of a
scalar field in an expanding universe.  We will see that the
dual projection procedure permit us to find a very different result from the
flat space one (section $II.A$). Obeying the results obtained by Fronsdal {\it et al} \cite%
{fronsdal}, we will show that the flat result can be obtained by the
application of a specific limit in the curved space outcome, as it should
be.  

{\bf A. A scalar field in a gravitational background}.
Before we approach the principal issue of this article, let us make an application of the dual projection to the case of a scalar field embedded in a general gravitational background.  In this way, we consider the following Lagrangian,
\be \l{a1}
{\cal L}_g\,=\, {1\over 2} \sqrt{-g}\,g^{\mu\nu}\,\p_\mu \phi \,\p_\nu \phi
\ee
where
\be
h^{\mu\nu}\,=\,\sqrt{-g}\,g^{\mu\nu}\,\p_\mu \phi \,\p_\nu \phi\,=\,
\left(
\begin{array}{cc}
h_{00} & h_{01} \\
h_{10} & h_{11}
\end{array}
\right) 
\ee
and $h_{01}=h_{10}$. 

Performing the decomposition in components of the action (\ref{a1}) we can write,
\be
{\cal L}_g\,=\, {1\over 2}\,h_{00}\,\dot{\phi}^2+\,{1\over 2}\,h_{11}\,{\phi ^{\prime }}%
^{2}\,+\,h_{01}\,\dot{\phi}\,\phi ^{\prime }
\ee

After the standard diagonalization process, as we explained in the last sections, we have as final result,
\be \l{a2}
{\cal L}\,=\,\varphi^{\prime}\,\dot{\varphi}\,-\,\eta_+\,{\varphi^{\prime}}%
^2\,-\,\sigma^{\prime}\,\dot{\sigma}\,-\, \eta_-\,{\sigma^{\prime}}^2\;\;,
\end{equation}
where $\eta_{\pm}={1 \over h_{00}}\,( 1\mp h_{01})$.  And (\ref{a2}) shows the coupling of the chiral bosons ($\varphi$ and $\sigma$) with the gravitational field.  This behavior is characteristic of a system composed by notons, which couple to the gravitational field \cite{aw}.

{\bf B. A scalar field in a curved expanding universe}.
Now we will analyze, at the light of the dual projection procedure, the
class of two-dimensional metrics of the form \cite{parker}, 
\begin{equation}  \label{1b}
ds^2\,=\,-\,dt^2\,+\,{R(t)}^2\,dx^2
\end{equation}
where $R(t)$ is an unspecified positive function of $t$. The two-dimensional
universe with such a metric will be referred conveniently, as we said
before, as an expanding universe. Consequently $R(t)$ need be increasing
with time. To work in two dimensions is, as it is well known, a laboratory
where it can be possible to envision the results in four dimensions or even
in higher dimensions.

The equations governing the fields are covariant generalizations of the
special relativistic free-field equation. The gravitational metric is
treated as an unquantized external field. No additional interactions are
included.

As a first step, let us verify if our space (\ref{1b}) is really a curved
one. It is well known that the Riemann-Christoffel tensor for
two-dimensional spaces has only nonvanishing component equal to $R_{0101}$
or to its negative. For the metric in (\ref{1b}), performing a standard
calculation of the $R_{0101}$ component, we have that, 
$
R_{0101}\,=\,-\,R\,\ddot{R}.
$
This gives us the condition that we are working in a two-dimensional curved
space only if $R\,\ddot{R} \ne 0$. On the other hand, for this space to be
flat, we would have to have the following condition, 
$
R_{0101}\,=\,0 \Longrightarrow  R\,\ddot{R}\,=\,0 ,
$
so that $R=0$ or $\ddot{R}\,=\,0$. The first solution, $R=0$ can be
obviously ignored. The second one, $\ddot{R}(t)=0$, has the straightforward
solution 
$
R(t)\,=\,k_1\,t\,+\,k_2\;\;,
$
where $k_1\,>\,0$ (to pledge an expansion of the universe) and $k_2$ are
arbitrary constants. So, as a condition for the space to be a curved one, $%
R(t)$ can not be a linear function of time. Unless for this case, we can say
that we are working in a curved space.

Back to the main issue, with the metric (\ref{1b}), the Lagrangian density for a two dimensional
scalar particle in a curved space is \cite{parker} 
\begin{equation}  \label{2b}
{\cal L}\,=\,{\frac{1 }{2}}\,R^3\,\left({\dot{\phi}}^2\,-\,{\frac{1 }{R^2}}\,%
{\phi^{\prime}}^2 \right)
\end{equation}
where we wrote for convenience that $R$ represents $R(t)$.

Now, let us perform the dual projection procedure following the steps of the
last section. Reducing the order of the time derivative in the same way as
we did in the last section, we can write 
\begin{equation}
{\frac{1}{2}}\,R^{3}\,{\dot{\phi}}^{2}\,\quad \rightarrow \quad \,p\,\dot{%
\phi}\,-\,{\frac{1}{2}}\,\frac{p^{2}}{R^{3}}\;\;.  \label{2.1}
\end{equation}
So, in (\ref{2b}) we have 
\begin{equation}
{\cal L}\,=\,p\,\dot{\phi}\,-\,{\frac{1}{2}}\,\frac{p^{2}}{R^{3}}%
\,-\,{\frac{1}{2}}\,R\,{\phi ^{\prime }}^{2}\;\;.  \label{3}
\end{equation}

Firstly, performing the generalized canonical transformations, 
\be  \label{41}
\phi = a\,\varphi \,+\,b\,\rho \qquad \mbox{and} \qquad
p =c\,\varphi ^{\prime }\,-\,d\,\rho ^{\prime }
\ee
where $a,b,c$ and $d$ are the coefficients that will be determined.
Substituting (\ref{41}) in (\ref{3}) we have 
\begin{eqnarray}  \label{5}
{\cal L} &=&a\,c\,\varphi ^{\prime }\dot{\varphi}\,-\,b\,d\rho ^{\prime }%
\dot{\rho}\,+\,(bc\,-\,ad)\phi ^{\prime }\,\dot{\rho}  \nonumber \\
&-&\,{\frac{1}{2}}\,\left( \,\frac{c^{2}}{R^{3}}\,+\,R\,a^{2}\right) \,{%
\varphi ^{\prime }}^{2}\,-\,{\frac{1}{2}}\,\left( \frac{d^{2}}{R^{3}}%
\,+\,R\,b^{2}\right) \,{\rho ^{\prime }}^{2}  \nonumber \\
&+&\left( \frac{cd}{R^{3}}\,-\,a\,bR\right) \,\varphi ^{\prime }\rho
^{\prime }\;\;.
\end{eqnarray}
Our first attempt is to obtain a flat space-type result so that the
cross-terms have to disappear. Hence, to compute the coefficients in (\ref{5}) we have,
\begin{equation}  \label{6.1}
ac\,=\,bd\,=\,1\quad \Longrightarrow \quad a={\frac{1}{c}}\qquad \mbox{and}%
\qquad b={\frac{1}{d}} \\
\end{equation}
\begin{equation}
bc\,-\,ad\,=\,0  \label{6.2} \\
\end{equation}
\begin{equation}
\mbox{and}\quad cd\,-\,ab\,R^{4}\,=\,0\;\;.  \label{6.4}
\end{equation}
We will call the coefficients of the fourth and fifth terms in (\ref{5}) as, 
\begin{eqnarray}  \label{7.1}
COEF_{1} &=&{\frac{1}{2}}\,\left( \frac{c^{2}}{R^{3}}\,+\,R\,a^{2}\right)\,=\,{\frac{1}{2}}\,\left( \frac{1}{a^{2}R^{3}}\,+\,R\,a^{2}\right) \nonumber \\
COEF_{2} &=&{\frac{1}{2}}\,\left( \frac{d^{2}}{R^{3}}\,+\,R\,b^{2}\right)\,=\,{\frac{1}{2}}\,\left( \frac{1}{b^{2}R^{3}}\,+\,R\,b^{2}\right)
,
\end{eqnarray}
where we used equations (\ref{6.1}).  Afterwards we will be back to them again.

Substituting the equation (\ref{6.1}) in (\ref{6.4})  we have,
\be
\label{7.2}
a^2\,b^2\,=\,{1 \over R^4}
\ee
and finally, using (\ref{6.1}) in (\ref{6.2}) we can write that,
\begin{equation}  \label{7}
a\,=\,\pm\,b  \qquad \mbox{and} \qquad c\,=\,\pm\,d
\end{equation}
and with these values it is easy to see that the coefficients $COEF_1$ and $%
COEF_2$ are identical, i.e.,
\begin{equation}  \label{metrica}
\eta\,=\,COEF_1\,=\,COEF_2\,=\,{\frac{1}{2}}\,\left( \frac{1}{a^2R^3}\,+\,%
R\,a^2 \right) \;\;.
\end{equation}

Solving equations (\ref{7.2}) and (\ref{7}) together, we conclude that exist four possible values for $a$ and $b$,
\be
\label{8}
a\,=\,\pm{1 \over R}\,,\,\pm{i \over R}\,=\,\pm\,b
\ee

So with the values obtained in (\ref{7}), and (\ref{8}) our canonical transformation (%
\ref{4}), are 
$
\phi=a\,(\,\varphi\,\pm\,\rho\,)$ and $
p=c\,(\,\varphi^{\prime}\,\mp\,\rho^{\prime}\,)$,
but we also know from (\ref{6.1}) that $a=1/c$, therefore we can finally
write, 
$
\phi={\frac{1}{c}}\,(\,\varphi\,\pm\,\rho\,)$ and 
$p=c\,(\,\varphi^{\prime}\,\mp\,\rho^{\prime}\,)$.
Making all the correspondent substitutions in (\ref{5}), the final
Lagrangian is 
\begin{equation} \l{9}
{\cal L}\,=\,\varphi^{\prime}\,\dot{\varphi}\,-\,\eta\,{\varphi^{\prime}}%
^2\,-\,\rho^{\prime}\,\dot{\rho}\,-\, \eta\,{\rho^{\prime}}^2\;\;,
\end{equation}
where
$$
\eta = {1 \over 2} \left( \frac{1}{a^2r^3}\,+\,r\,a^2 \right)\;\;.
$$
Also, using (\ref{8}), we can have two values for $\eta$,
$
\eta_+\,=\,{1 \over R}
$
and 
$
\eta_-\,=\,-\,{1 \over R}.
$

We can see from (\ref{9}), as we said before, that $\varphi,\rho$  are two dynamical chiral fields interacting with the external gravitational field.  This result differs altogether from that found in the previous section, where we found two notons. There we can be naively convinced, at a first sight, that the final result is independent from the form of the gravitational background.  We see now that it is not totally true.    These two chiral fields coupling, in fact, describes a single dynamical non-chiral field in the given gravitational background, which was our starting point.  This will be confirmed just below.

To corroborate the condition that the flat space result is a limit of the
curved space one \cite{fronsdal}, let us fix the value of $\eta$ as 
$
\eta\,=\,1
$
and hence from (\ref{metrica}) $COEF_1\,=\,COEF_2=1$, originating conditions on $a$, 
\begin{equation}
a^4\,-\,{2\over R}\,a^2\,+\,{1\over R^4}\,=\,0
\end{equation}
which solutions are 
\begin{eqnarray}  
a_{1,2}&=&\pm\,{1\over R^2}\sqrt{R\,+\,\sqrt{R^2-1}}  \nonumber \\
a_{3,4}&=&\pm\,{1\over R^2}\sqrt{R\,-\,\sqrt{R^2-1}}\;\;.
\end{eqnarray}
Applying simply $\eta=1$ in (\ref{9}) we have the same Lagragian of the flat space, 
\begin{equation}
{\cal L}\,=\,\varphi^{\prime}\,\dot{\varphi}\,-\,{\varphi^{\prime}}%
^2\,-\,\rho^{\prime}\,\dot{\rho}\,-\,{\rho^{\prime}}^2\;\;,
\end{equation}
i.e., it comprises two FJ's particles with opposite chiralities, as stressed
in \cite{tw,adw} and confirms the existence of two chiral fields interacting with an external gravitational background as the result above in (\ref{9}).

\section{Conclusions}

The dual projection procedure is a technique that helps one to analyze the spectrum
of first-order systems.  It was used recently to study the notion of
duality and self-duality creating an internal space of potentials \cite{baw}.

Since the analysis was always effected for first-order systems, an
equivalence between the Lagrangian and Hamiltonian approaches permitted us
to use the concept of canonical transformations. In other words we can say
that the dual projection demanded a change of variables which was, in the
phase space, a canonical transformation.

In this work we consider the structure of a scalar field in a
two-dimensional expanding universe embedded in a curved space. 
Firstly, we construct a scalar field embedded in a generic gravitational background and promote the dual projection of the system.  We obtain, as expected, two notons.  Differently
from the generic case, now in a curved space, we have two dynamical chiral fields interacting with an external gravitational field.  This is equivalent to the model describing a single dynamical non-chiral field in the gravitational background.  
In other words, only the final models are equivalent, the spectra ($noton_1 \oplus noton_2$ and $chiral \:particle_1 \oplus chiral\: particle_2$) are not.  This final analysis is the main idea of this work.

However, as we can see, we could expect (naively based on the former result) that every other case involving this kind of coupling, i.e., whatever the form of the gravitational background, we always would find two notons comprising the spectrum.  The only difference between the spectra would manifest itself in the explicit form of the metric.

We show here that any conclusion about the spectrum turns out to be premature and thereby must be preceded by a detailed (for example) dual projection analysis.  In this way we can determine precisely if the particles that comprise the spectrum can be either notons or chiral fields coupled to the gravitational background although the final results are the same (a scalar field) as said above.
 

\section{Acknowledgments}

The author would like to thank professor Maur\'{\i}cio Calv\~{a}o for
valuable discussions and Funda\c{c}\~ao de Amparo \`a Pesquisa do Estado do
Rio de Janeiro (FAPERJ) for financial support.


\end{document}